\documentclass[onecolumn,showpacs,preprintnumbers,amsmath,amssymb]{revtex4}
\usepackage{graphicx}
\usepackage{dcolumn}
\usepackage{bm}

\begin{document}

\title{Best Fidelity Conditions for Three Party Quantum Teleportation}
\author{Sangjin Sim, Jeongwan Jin}
\author{Younghun Kwon}
\email{yyhkwon@hanyang.ac.kr}
\affiliation{Department of Physics, Hanyang University, \\Ansan, Kyunggi-Do, 425-791, South Korea\\}

\date{\today}
\begin{abstract}
Recently Acin et al. classified three qubit states and showed that there would be various types of three qubit entangled states. Using the entangled three qubit states classified by Acin et al. we consider the quantum teleportation among three parties. And we find the best fidelity conditions for the quantum teleportation among three parties.
\end{abstract}
\maketitle

\vspace{-.4in}
Quantum teleportation is an efficient  way to transmit a quantum information when classical communications(CC) and local operations(LO) are  allowed to parties already sharing an entangled state. The quantum teleportation between two parties was proposed by Bennet et al\cite{1}. In quantum teleportation of two parties, Alice(sender) and Bob(receiver) share a maximally entangled state. Alice attaches the information state to shared entangled state and performs Bell basis mesurement. And she sends her result to Bob. According to Alice's Bell basis measurement, Bob applies the corresponding unitary operation on his single qubit and  obtains the original information state with certainty.\\

\par Quantum teleportation of three parties using a three qubit entangled state was introduced by Karsson et al\cite{2}. The main difference between quantum teleportation of the two parties and that of three parties is the concept of cosender. A sender first performs Bell basis measurement on his(or her) two qubits,(one is the information qubit and the other is the qubit entangled to other parties) and sends the measurement result to the co-sender and the receiver.The co-sender performs single qubit measurement according to the sender's measurement result and sends the measurement result to the receiver.Given the protocol provided in the secret when three parties are separated, the receiver performs local unitary operations according to the measurement results. Then the receiver can recover the information state with a probability. \\
Recently Acin et al. showed that there are 7 different entangled states in three qubit states\cite{4}. In this paper, we consider quantum teleportation in three parties with those entangled states.In fact Yeo considered the quantum teleportation among three parties, using GHZ state and W state\cite{3}. So we will consider the quantum teleportation among three parties, using the other entangled states except GHZ state and W state. For each case, we will provide the fidelity, the best fidelity condition and teleportation protocols. 

\par This paper is organized as  
follows. In the section I, we first review the three party quantum teleportation with GHZ state and W state. In the section II, we consider quantum teleportation in three parties sharing different entangled states based on Acin et al.'s classification of three qubit states. And the roles of parties, sender, co-sender, and receiver, are determined. Also we give the best fidelity conditions for each case.
In the section III, we summarize and discuss our results. 
\section*{I. QUANTUM TELEPORTATION IN THREE PARTIES WITH SYMMETRIC THREE-QUBIT STATES }

Let us review quantum teleportation among three parties. Quantum teleportation in 
three parties sharing a three-qubit entangled state consists of three steps: \\

1)First, three parties shares a three qubit entangled state. A sender performs Bell basis measurement on his(or her) two qubits,(one is the information qubit and the other the qubit  
entangled to other parties) and sends the measurement result $j$ to the co-sender and the receiver. The Bell basis measurement  
makes use of the following projection operators : $|\Phi^{+}\rangle \langle \Phi^{+}|$ for $j=1$, $|\Phi^{-}\rangle \langle  
\Phi^{-}|$ for $j=2$, $|\Psi^{+}\rangle \langle \Psi^{+}|$ for $j=3$, and $|\Psi^{-}\rangle \langle \Psi^{-}|$ for $j=4$,  
where

\begin{eqnarray}
|\Phi^{\pm}\rangle & = & \frac{1}{\sqrt{2}}(|00\rangle \pm |11\rangle) \\
|\Psi^{\pm}\rangle & = & \frac{1}{\sqrt{2}}(|01\rangle \pm |10\rangle) \\ \nonumber\\
\end{eqnarray}

2)The co-sender performs single qubit measurement, according to the sender's measurement result $j$, and sends the  
measurement result $k$ to the receiver. The single qubit measurement applies the following pojections : $|\mu^{+}\rangle \langle  
\mu^{+}|$ for $k=1$ and $|\mu^{-}\rangle \langle \mu^{-}|$ for $k=2$, where

\begin{eqnarray}
|\mu^{+}\rangle & = & \sin\nu|0 \rangle + e^{i\kappa}\cos\nu|1 \rangle \\
|\mu^{-}\rangle & = & \cos\nu|0 \rangle - e^{i\kappa}\sin\nu|1 \rangle \\ \nonumber\\
\end{eqnarray}

3)Given the protocol provided in the secret when three parties are separated, the receiver performs local unitary operations according  
to the measurement results $j$ and $k$. Then the party recovers the information state with a probability. \\

If $|\tau \rangle$ denotes the receiver's reconstructed state, then the success rate is measured by the fidelity of the original information state,  
$|\psi\rangle = \cos(\theta/2)|0\rangle + e^{i\phi}\sin(\theta/2)|1\rangle$, and the $|\tau\rangle$, which is 

\begin{eqnarray}
\langle F\rangle  = \frac{1}{4\pi}\int_{0}^{2\pi}d\phi \int_{0}^{\pi} \sin\theta d\theta \sum_{j,k}|\langle  
\psi|\tau\rangle|^{2} \nonumber \\
\end{eqnarray}

When the GHZ state is shared in three parties, the fidelity is shown as $\langle F_{GHZ}\rangle = \frac{2} 
{3}+\frac{1}{3}\sin2\nu$ given the protocol in table I. \\

\begin{table}[here]
\begin{center}
\begin{tabular}{cccccc}
    &  & $j=1$ & $j=2$ & $j=3$ & $j=4$  \\
\hline \\
$k=1$ & \vline  & $I$ & $\sigma_{z}$ & $\sigma_{x}$ & $\sigma_{y}$ \\
$k=2$ & \vline  & $\sigma_{z}$ & $I$ & $\sigma_{y}$ & $\sigma_{x}$ \\
\end{tabular}
\caption{\label{I}
The protocol of the quantum teleportation in three parties when the GHZ states is applied} 
\end{center}
\end{table}

When the W state is shared in three parties, the fidelity is shown as $\langle F_{W}\rangle = 7/9$ given the protocol  in table II.

\begin{table}[here]
\begin{center}
\begin{tabular}{cccccc}
    &  & $j=1$ & $j=2$ & $j=3$ & $j=4$  \\
\hline \\
$k=1$ & \vline  & $\sigma_{x}$ & $\sigma_{y}$ & $I$ & $\sigma_{z}$ \\
$k=2$ & \vline  & $\sigma_{x}$ & $\sigma_{y}$ & $I$ & $\sigma_{z}$ \\
\end{tabular}
\caption{\label{II}
The protocol of the quantum teleportation in three parties when the W states is applied} 
\end{center}
\end{table}

\newpage
We here note that $\langle F_{W}\rangle > \langle F_{GHZ}\rangle$ in average. However if $sin2\nu$ is greater than $\frac{1}{3}$, $F_{GHZ} > F_{W}$. And the best fidelity condition for $F_{GHZ}$ is $\nu = \frac{\pi}{4}+m\pi$. Here the best fidelity condition means that if the co-sender Bob can perform his single qubit measurement, according to the sender's measurement result $j$, using the following pojections : $|\mu^{+}\rangle \langle  
\mu^{+}|$ for $k=1$ and $|\mu^{-}\rangle \langle \mu^{-}|$ for $k=2$, where

\begin{eqnarray}
|\mu^{+}\rangle & = & \frac{1}{\sqrt{2}} |0 \rangle + \frac{1}{\sqrt{2}} |1 \rangle \\
|\mu^{-}\rangle & = & \frac{1}{\sqrt{2}}|0 \rangle - \frac{1}{\sqrt{2}}|1 \rangle \\ \nonumber\\
\end{eqnarray}\\

then the fidelity produces the best result.\\

\section*{II. Quantum teleportation with asymmetric states}

The classification of three-qubit state by Acin et al. is as follows;

\par Type 1 (Product states)\\

\par Type 2a (Biseparable states)\\

\begin{eqnarray}
|2aI\rangle & = & \frac{1}{\sqrt{3}}(|000\rangle + |100\rangle + |101\rangle) \nonumber \\
|2aII\rangle & = &  \frac{1}{\sqrt{3}}(|000\rangle + |100\rangle + |110\rangle) \nonumber \\
\end{eqnarray}

\par Type 2b (GHZ state)\\

\begin{eqnarray}
|2b\rangle & = & \frac{1}{\sqrt{2}}(|000\rangle + |111\rangle ) \nonumber \\
\end{eqnarray}

\par Type 3a (Tri-Bell state)

\begin{eqnarray}
|3a\rangle & = & \frac{1}{\sqrt{3}}(|000\rangle + |101\rangle + |110\rangle) \nonumber  \\
\end{eqnarray}

\par Type 3b (Extended GHZ states)

\begin{eqnarray}
|3bI\rangle & = & \frac{1}{\sqrt{3}}(|000\rangle + |110\rangle + |111\rangle) \nonumber \\
|3bII\rangle & = & \frac{1}{\sqrt{3}}(|000\rangle + |100\rangle + |111\rangle) \nonumber \\
|3bIII\rangle & = & \frac{1}{\sqrt{3}}(|000\rangle + |101\rangle + |111\rangle) \nonumber \\
\end{eqnarray}

\par Type 4a 

\begin{eqnarray}
|4a\rangle & = & \frac{1}{\sqrt{4}}(|000\rangle + |100\rangle + |101\rangle + |110\rangle) \nonumber \\
\end{eqnarray}

\par Type 4b 

\begin{eqnarray}
|4bI\rangle & = & \frac{1}{\sqrt{4}}(|000\rangle + |100\rangle + |110\rangle + |111\rangle) \nonumber \\
|4bII\rangle & = & \frac{1}{\sqrt{4}}(|000\rangle + |100\rangle + |101\rangle + |111\rangle) \nonumber \\
\end{eqnarray}

\par Type 4c 

\begin{eqnarray}
|4c\rangle & = & \frac{1}{\sqrt{4}}(|000\rangle + |101\rangle + |110\rangle + |111\rangle) \nonumber \\
\end{eqnarray}

\par Type 5 (Real states)

\begin{eqnarray}
|5\rangle & = & \frac{1}{\sqrt{5}}(|000\rangle + |100\rangle + |101\rangle + |110\rangle + |111\rangle) \nonumber \\
\end{eqnarray}

Note that the tri-Bell state is equivalent to the W state. Therefore, we need to consider only type3-5 states.
\par We now show all schemes of quantum teleportation in three parties sharing one of the these types. Note the reference of  
protocols. The two W protocols are equivalent with respect to a permutation of parties. 

\begin{table}[here]
\begin{center}
\begin{tabular}{cccccc}
    &  & $j=1$ & $j=2$ & $j=3$ & $j=4$  \\
\hline \\
$k=1$ & \vline  & $I$ & $\sigma_{z}$ & $\sigma_{x}$ & $\sigma_{y}$ \\
$k=2$ & \vline  & $\sigma_{z}$ & $I$ & $\sigma_{y}$ & $\sigma_{x}$ \\
\end{tabular}
\caption{GHZ protocol}
\end{center}
\end{table}

\begin{table}[here]
\begin{center}
\begin{tabular}{cccccc}
    &  & $j=1$ & $j=2$ & $j=3$ & $j=4$  \\
\hline \\
$k=1$ & \vline  & $I$ & $\sigma_{z}$ & $\sigma_{x}$ & $\sigma_{y}$ \\
$k=2$ & \vline  & $I$ & $\sigma_{z}$ & $\sigma_{x}$ & $\sigma_{y}$\\
\end{tabular}
\caption{W protocol I}
\end{center}
\end{table}

\begin{table}[here]
\begin{center}
\begin{tabular}{cccccc}
    &  & $j=1$ & $j=2$ & $j=3$ & $j=4$  \\
\hline \\
$k=1$ & \vline  & $\sigma_{x}$ & $\sigma_{y}$ & $I$ & $\sigma_{z}$\\
$k=2$ & \vline  & $\sigma_{x}$ & $\sigma_{y}$ & $I$ & $\sigma_{z}$\\
\end{tabular}
\caption{W protocol II}
\end{center}
\end{table}

\subsection{Quantum teleportation in three parties sharing the extended GHZ state}

An extended GHZ state is transformed to another extended GHZ state under permutations of parties. Thus, it is sufficient to consider  
the case that Alice, Bob and Cindy share the following state

\begin{eqnarray}
|\psi \rangle & = & \frac{1}{\sqrt{3}}(|0_{A}0_{B}0_{C}\rangle + |0_{A}1_{B}1_{C}\rangle + |1_{A}1_{B}1_{C}\rangle)  
\nonumber \\
\end{eqnarray}

Suppose that they want to teleport the information state $\cos(\theta/2)|0\rangle + e^{i\kappa}\sin(\theta/2)|1\rangle$. We  
know that Bob and Cindy are symmetric to a permutation of them. There are four choices in determining their roles of quantum  
teleportation. We will use '$\rightarrow$' to mean that a party sends the measurement result to another one via CC(one-way) and  
'$\leftrightarrow$' to mean that both $\rightarrow$ and $\leftarrow$ are possible via CC(two-way). \\

1. Alice(sender) $\rightarrow$ Bob $\leftrightarrow$ Cindy \\

The fidelity is $\frac{5}{9}+\frac{2}{9}\cos\kappa\sin2\nu$ and the protocol is of GHZ. The best fidelity condition is $\kappa=2n\pi, \nu=\frac{\pi}{4}+m\pi$\\

2. Bob(sender) $\rightarrow$ Alice(co-sender) $\rightarrow$ Cindy(receiver) \\

The fidelity is $\frac{8}{9}$, and the protocol is of W. \\
 
3. Bob(sender) $\rightarrow$ Cindy(co-sender) $\rightarrow$ Alice(receiver) \\

The fidelity is $\frac{5}{9}+\frac{2}{9}\cos\kappa\sin2\nu$, and the protocol is of GHZ. The best fidelity condition is $\kappa=2n\pi, \nu=\frac{\pi}{4}+m\pi$\\\\

\subsection{Quantum teleportation in three parties sharing the type4a state}

Suppose that Alice, Bob and Cindy shared the following state

\begin{eqnarray}
|\psi \rangle & = & \frac{1}{\sqrt{4}}(|0_{A}0_{B}0_{C}\rangle + |1_{A}0_{B}0_{C}\rangle + |1_{A}0_{B}1_{C}\rangle  
\nonumber \\ &&+|1_{A}1_{B}0_{C}\rangle) \nonumber \\
\end{eqnarray}

Since Bob and Cindy are symmetric parties, there are four cases as follows, \\

1. Alice(sender) $\rightarrow$ Bob $\leftrightarrow$ Cindy \\

The fidelity is $\frac{2}{3}$ and the protocol is of W. \\

2.  Bob(sender) $\rightarrow$ Alice(co-sender) $\rightarrow$ Cindy(receiver) \\

The fidelity is $\frac{2}{3}$ and the protocol is of the second W. \\

3. Bob(sender) $\rightarrow$ Cindy(co-sender) $\rightarrow$ Alice(receiver) \\

The fidelity is $\frac{2}{3}$ and the protocol is of W. \\

\subsection{Quantum teleportation in three parties sharing the type4b state}

Suppose that Alice, Bob and Cindy shared the following state

\begin{eqnarray}
|\psi \rangle & = & \frac{1}{\sqrt{4}}(|0_{A}0_{B}0_{C}\rangle + |1_{A}0_{B}0_{C}\rangle + |1_{A}1_{B}0_{C}\rangle  
\nonumber \\ && +|1_{A}1_{B}1_{C}\rangle) \nonumber \\
\end{eqnarray}

Since there are no symmetric parties, there are six cases as follows, \\

1. Alice(sender) $\rightarrow$ Bob(co-sender) $\leftrightarrow$ Cindy(receiver) \\

The fidelity is $\frac{1}{2}+\frac{1}{6}\cos\kappa \sin2\nu$ and the protocol is of GHZ. The best fidelity condition is $\kappa=2n\pi, \nu=\frac{\pi}{4}+m\pi$\\\\

2. Alice(sender) $\rightarrow$ Cindy(co-sender) $\rightarrow$ Bob(receiver) \\

The fidelity is $\frac{3}{4}$ and the protocol is of W. \\

3. Bob(sender) $\rightarrow$ Alice $\rightarrow$ Cindy \\

The fidelity is $\frac{7}{12}+\frac{1}{6}\cos2\nu+\frac{1}{6}\cos\kappa \sin2\nu$ and the protocol is of GHZ. The best fidelity condition is $\kappa=2n\pi, \nu=\frac{\pi}{8}+m\pi$\\\\

4. Bob(sender) $\rightarrow$ Cindy(co-sender) $\rightarrow$ Alice(receiver) \\

The fidelity is $\frac{3}{4}$ and the protocol is of W. \\

5. Cindy(sender) $\rightarrow$ Alice(co-sender) $\rightarrow$ Bob(receiver) \\

The fidelity is $\frac{7}{12}+\frac{1}{6}\cos2\nu+\frac{1}{6}\cos\kappa \sin2\nu$ and the protocol is of GHZ. The best fidelity condition is $\kappa=2n\pi, \nu=\frac{\pi}{8}+m\pi$\\\\

6. Cindy(sender) $\rightarrow$ Bob(co-sender) $\rightarrow$ Alice(receiver) \\

The fidelity is $\frac{1}{2}+\frac{1}{6}\cos\kappa\sin2\nu$ and the protocol is of GHZ. The best fidelity condition is $\kappa=2n\pi, \nu=\frac{\pi}{4}+m\pi$\\\\

\subsection{Quantum teleportation in three parties sharing the type4c state}

Suppose that Alice, Bob and Cindy shared the following state

\begin{eqnarray}
|\psi \rangle & = & \frac{1}{\sqrt{4}}(|0_{A}0_{B}0_{C}\rangle + |1_{A}0_{B}1_{C}\rangle + |1_{A}1_{B}0_{C}\rangle  
\nonumber \\ &&+|1_{A}1_{B}1_{C}\rangle) \nonumber \\
\end{eqnarray}

Since Bob and Cindy are symmetric parties, there are four cases as follows, \\

1. Alice(sender) $\rightarrow$ Bob $\leftrightarrow$ Cindy \\

The fidelity is $\frac{3}{4}$ and the protocol is of W. \\

2. Bob(sender) $\rightarrow$ Alice(co-sender) $\rightarrow$ Cindy(receiver) \\

The fidelity is $\frac{1}{2}+\frac{1}{6}\cos\kappa \sin2\nu$ and the protocol is of the second GHZ. The best fidelity condition is $\kappa=2n\pi, \nu=\frac{\pi}{4}+m\pi$\\\\

3. Bob(sender) $\rightarrow$ Cindy(co-sender) $\rightarrow$ Alice(receiver) \\

The fidelity is $\frac{3}{4}$ and the protocol is of W.\\

\subsection{Quantum teleportation in three parties sharing the type5 state}

Suppose that Alice, Bob and Cindy shared the following state

\begin{eqnarray}
|\psi \rangle & = & \frac{1}{\sqrt{4}}(|0_{A}0_{B}0_{C}\rangle + |1_{A}0_{B}0_{C}\rangle + |1_{A}0_{B}1_{C}\rangle  
\nonumber \\ && + |1_{A}1_{B}0_{C}\rangle+|1_{A}1_{B}1_{C}\rangle) \nonumber \\
\end{eqnarray}

Since Bob and Cindy are symmetric parties, there are four cases as follows, \\

1. Alice(sender) $\rightarrow$ Bob $\leftrightarrow$ Cindy \\

The fidelity is $\frac{2}{3}$ and the protocol is of W. \\

2. Bob(sender) $\rightarrow$ Alice(co-sender) $\rightarrow$ Cindy(receiver) \\

The fidelity is $\frac{8}{15}+\frac{2}{15}\cos2\nu + \frac{2}{15}\cos\kappa \sin 2\nu$ and the protocol is of GHZ. The best fidelity condition is $\kappa=2n\pi, \nu=\frac{\pi}{8}+m\pi$\\\\

3. Bob(sender) $\rightarrow$ Cindy(co-sender) $\rightarrow$ Alice(receiver) \\

The fidelity is $\frac{2}{3}$ and the protocol is of W. \\
All scheme of quantum teleportatopn among three parties are shown in Table VI.

\begin{table}[here]
\begin{center}
\begin{tabular}{cccccc}
    &  & $role$ & $fidelity$ & $protocol$ & $best fidelity condition$  \\
\hline \\
$extended GHZ$ & \vline  & Alice(sender) $\rightarrow$ Bob $\leftrightarrow$ Cindy & $GHZ$ & $\frac{5}{9}+\frac{2}{9}\cos\kappa\sin2\nu$ & $\kappa=2n\pi, \nu=\frac{\pi}{4}+m\pi$ \\
 & \vline  & Bob(sender) $\rightarrow$ Alice(co-sender) $\rightarrow$ Cindy(receiver) & $W$ & $\frac{8}{9}$ &  \\
 & \vline  & Bob(sender) $\rightarrow$ Cindy(co-sender) $\rightarrow$ Alice(receiver) & $GHZ$ & $\frac{5}{9}+\frac{2}{9}\cos\kappa\sin2\nu$ & $\kappa=2n\pi, \nu=\frac{\pi}{4}+m\pi$ \\
$type 4a$ & \vline  & Alice(sender) $\rightarrow$ Bob $\leftrightarrow$ Cindy & $W$ & $\frac{2}{3}$ & \\
 & \vline  & Bob(sender) $\rightarrow$ Alice(co-sender) $\rightarrow$ Cindy(receiver) & $W$ & $\frac{2}{3}$ &  \\
 & \vline  & Bob(sender) $\rightarrow$ Cindy(co-sender) $\rightarrow$ Alice(receiver) & $W$ & $\frac{2}{3}$ &  \\
$type 4b$ & \vline  & Alice(sender) $\rightarrow$ Bob(co-sender) $\leftrightarrow$ Cindy(receiver) & $GHZ$ & $\frac{1}{2}+\frac{1}{6}\cos\kappa \sin2\nu$ & $\kappa=2n\pi, \nu=\frac{\pi}{4}+m\pi$ \\
 & \vline  & Alice(sender) $\rightarrow$ Cindy(co-sender) $\rightarrow$ Bob(receiver) & $W$ & $\frac{3}{4}$ &  \\
 & \vline  & Bob(sender) $\rightarrow$ Alice $\rightarrow$ Cindy & $GHZ$ & $\frac{7}{12}+\frac{1}{6}\cos2\nu+\frac{1}{6}\cos\kappa \sin2\nu$ & $\kappa=2n\pi, \nu=\frac{\pi}{8}+m\pi$ \\
 & \vline  & Bob(sender) $\rightarrow$ Cindy(co-sender) $\rightarrow$ Alice(receiver) & $W$ & $\frac{3}{4}$ &  \\
 & \vline  & Cindy(sender) $\rightarrow$ Alice(co-sender) $\rightarrow$ Bob(receiver) & $GHZ$ & $\frac{7}{12}+\frac{1}{6}\cos2\nu+\frac{1}{6}\cos\kappa \sin2\nu$ & $\kappa=2n\pi, \nu=\frac{\pi}{8}+m\pi$ \\
 & \vline  & Cindy(sender) $\rightarrow$ Bob(co-sender) $\rightarrow$ Alice(receiver) & $GHZ$ & $\frac{1}{2}+\frac{1}{6}\cos\kappa\sin2\nu$ & $\kappa=2n\pi, \nu=\frac{\pi}{4}+m\pi$ \\
$type 4c$ & \vline  & Alice(sender) $\rightarrow$ Bob $\leftrightarrow$ Cindy & $W$ & $\frac{3}{4}$ &  \\
 & \vline  & Bob(sender) $\rightarrow$ Alice(co-sender) $\rightarrow$ Cindy(receiver) & $GHZ$ & $\frac{1}{2}+\frac{1}{6}\cos\kappa \sin2\nu$ & $\kappa=2n\pi, \nu=\frac{\pi}{4}+m\pi$ \\
 & \vline  & Bob(sender) $\rightarrow$ Cindy(co-sender) $\rightarrow$ Alice(receiver) & $W$ & $\frac{3}{4}$ & \\
$type 5$ & \vline  & Alice(sender) $\rightarrow$ Bob $\leftrightarrow$ Cindy & $W$ & $\frac{2}{3}$ & \\
 & \vline  & Bob(sender) $\rightarrow$ Alice(co-sender) $\rightarrow$ Cindy(receiver) & $GHZ$ & $\frac{8}{15}+\frac{2}{15}\cos2\nu + \frac{2}{15}\cos\kappa \sin 2\nu$  & $\kappa=2n\pi, \nu=\frac{\pi}{8}+m\pi$ \\
 & \vline  & Bob(sender) $\rightarrow$ Cindy(co-sender) $\rightarrow$ Alice(receiver) & $W$ & $\frac{2}{3}$ & \\

\end{tabular}
\caption{\label{VI}
table for three party quantum teleportation} 
\end{center}
\end{table}

\par  We here note that there are only two protocols W and GHZ in tables[III] and [IV]. This is due to the different entanglement structure of W and GHZ states. In  
other words, W state cannot be transformed to GHZ state with a probability. This implies that protocols can classify quantum  
states like stochastic LOCC. That is, we classify the five three-qubit states to two classes, W  
and GHZ, based on protocols.

\begin{table}[here]
\begin{center}
\begin{tabular}{cccccc}

      & GHZ-type &  & W-type &   \\
\hline \\
& Type2b  &  \vline & Type3, Type4, and Type5  &  \\
\end{tabular}
\end{center}
\end{table}

\section*{III. CONCLUDING REMARK}
In this report, we provided all schemes of quantum teleportation in three parties. Referring Acin et al.'s classification, which is based on the  
Schmidt decompostion with the set of strongly asymmetric basis, we considered all cases of quantum teleportation in three parties. We obtained the best fidelity condition for each case. We also assigned the the roles(sender, co-sender, and receiver) of the parties generically.

\section*{Acknowledgement}
Y. Kwon is supported by the Fund of Hanyang University.

\end{document}